\documentstyle[12pt]{article}
\newcommand{\be}{\begin{equation}}
\newcommand{\ee}{\end{equation}}
\newcommand{\la}{\lambda}
\newcommand{\al}{\alpha}
\newcommand{\La}{\Lambda}
\newcommand{\p}{\partial}

\newcommand{\ba}{\begin{array}}
\newcommand{\ea}{\end{array}}
\newcommand{\lb}{\label}

\begin{document}
\setlength{\baselineskip}{6mm}
\setlength{\textwidth}{160mm}
\setlength{\textheight}{230mm}
\setlength{\topmargin}{-1cm}
\renewcommand{\arraystretch}{1.3}

\title{\  \\ \  \\Classical Poisson structures and $r$-matrices
\\from constrained flows}
\author{ Yunbo Zeng${}^{1,2}$ and Jarmo Hietarinta${}^{1}$ \\
${}^1$Department of Physical Sciences, University of Turku\\
FIN-20500 Turku, Finland \\
${}^2$Department of Applied Mathematics,Tsinghua University\\
Beijing 100084, China\thanks{permanent address of Yunbo Zeng.}}
\maketitle
\begin{abstract} We construct the classical Poisson structure and
$r$-matrix for some finite dimensional integrable Hamiltonian systems
obtained by constraining the flows of soliton equations in a certain
way. This approach allows one to produce new kinds of classical, dynamical
Yang-Baxter structures.  To illustrate the method we present the
$r$-matrices associated with the constrained flows of the Kaup-Newell,
KdV, AKNS, WKI and TG hierarchies, all generated by a 2-dimensional
eigenvalue problem.  Some of the obtained $r$-matrices depend only on
the spectral parameters, but others depend also on the dynamical
variables.  For consistency they have to obey a classical
Yang-Baxter-type equations, possibly with dynamical extra terms.
\end{abstract}

\newpage
\section {Introduction}
Integrable finite dimensional systems that admit a classical $r$-matrix
depending only on the spectral parameters has been studied extensively
\cite{l86}.  Recently it has been found that for many integrable systems the
$r$-matrix depends also on the dynamical variables
\cite{o90,jm85,ms83,jc94,skl,ja93}.  For example, the celebrated
Calogero-Moser system has been shown to possess a dynamical $r$-matrix
\cite{skl,ja93}. In contrast with the well-studied case of
$r$-matrices depending only on spectral parameters, the general theory
of dynamical $r$-matrices has not yet been established. New examples
of dynamical $r$-matrices are therefore needed for the search for the
underlying structure, and the method presented below seems to be quite
useful for this purpose.

In recent years, many types constrained flows of soliton hierarchies
have been discussed in the literature. For one such class the Lax
representation can be deduced from the adjoint representation of the
auxiliary linear problems of the soliton equation \cite{z93a,z94}, or
derived by using the Gelfand-Dikii approach \cite{m93}. By means of
the Lax representation one can then construct the classical Poisson
structure and $r$-matrix for the constrained flows. For some
constrained flows the $r$-matrix depends only on the spectral
parameters, but for others it also depends on the dynamical variables.

For consistency the Poisson bracket has to obey the Jacobi identity
and this implies an equation for the $r$-matrix. In some cases this
equation is just the classical Yang-Baxter equation, but in other
cases there will be dynamical extra terms.

In present paper, to illustrate the above, we describe the classical
Poisson structure and the related classical Yang-Baxter equations
associated with the constrained flows for the Kaup-Newell hierarchy, the
AKNS hierarchy and the KdV hierarchy. We present also two examples of
dynamical $r$-matrix associated with the G. Tu (TG) hierarchy and the
Wadati-Konno-Ichikawa (WKI) hierarchy and discuss the related
``classical, dynamical'' Yang-Baxter equation. At same time some new
solutions of the dynamical Yang-Baxter equations are obtained.

\section {Integrable constrained flows}
To make the paper self contained we first briefly describe how finite
dimensional integrable systems and their Lax representation can be
constructed from constrained flows of soliton equations. We will use
the Kaup-Newell (KN) hierarchy as an illustration, for further details, see
\cite{z94}.

\subsection{The hierarchy of Hamiltonian flows}
Let us start by considering the the Kaup-Newell eigenvalue problem
\cite{kn78}
\be
\left( \begin{array}{c}\psi_{1}\\\psi_{2}\end{array}\right)_{x}=
U(u, \lambda)
\left( \begin{array}{c}\psi_{1}\\\psi_{2}\end{array}\right),
\quad U(u, \lambda)
=\left( \begin{array}{cc}-\lambda^{2}&\lambda q\\\lambda r&
\lambda^{2}\end{array}\right).\label{a1}
\ee
[Here and in the following we denote $u^t=(q,r)$.]
First, we solve the adjoint representation of (\ref{a1}) \cite{gu89,ac85}
\be
V_{x}=[U,V]\equiv UV-VU,\label{d1}
\ee
where $V$ has a Laurent series expansion
\be
V(u,\la)=\sum_{m=0}^{\infty}\left( \begin{array}{cc}a_{m}(u)&
b_{m}(u)\\c_{m}(u)&-a_{m}(u)
\end{array}\right)\la^{-m}. \label{d2}
\ee
Eqs (\ref{d1}) and (\ref{d2}) lead to the recursion relations
\be\ba{l}
b_{m+2}=-qa_{m+1}-\frac {1}{2}b_{m,x},\\
c_{m+2}=-ra_{m+1}+\frac {1}{2}c_{m,x},\\
a_{m}=\frac {1}{2}\partial_{x}^{-1}(qc_{m-1,x}+rb_{m-1,x}),
\ea\lb{d3}\ee
and to the parity constraints $a_{2m+1}=b_{2m}=c_{2m}=0$. The first
few terms are as follows:
\be\ba{l}
a_{0}=1,\quad a_{2}=-\frac {1}{2}qr,\quad
a_{4}=\frac {3}{8}q^{2}r^{2}+\frac {1}{4}(rq_{x}-qr_{x}),\dots\\
b_{1}=-q,\quad b_{3}=\frac {1}{2}(q^{2}r+q_{x}),\dots\\
c_{1}=-r,\quad c_{3}=\frac {1}{2}(r^{2}q-r_{x}),\cdots
\ea\lb{d3x}\ee
The recursion relation (\ref{d3}) can also be expressed as
\be
\left (\begin{array}{c}c_{2m+1}\\b_{2m+1}\end{array}\right)
=L\left (\begin{array}{c}c_{2m-1}\\b_{2m-1}\end{array}\right),\quad
L=\frac{1}{2}\left( \begin{array}{cc}\partial_{x}-
r\partial_{x}^{-1}q\partial_{x}&-r\partial_{x}^{-1}
r\partial_{x}\\-q\partial_{x}^{-1}q\partial_{x}&
-\partial_{x}-q\partial_{x}^{-1}r\partial_{x}\end{array}\right).
\label{d4}
\ee

Next let us consider a ``truncation'' of the expression (\ref{d2})
\be
V^{(n)}(u,\la)\equiv (\la^{2n} V)_+ \equiv
\sum_{m=0}^{n-1}\left( \begin{array}{cc}a_{2m}\lambda^{2n-2m}&
b_{2m+1}\lambda^{2n-2m-1}\\
c_{2m+1}\lambda^{2n-2m-1}&
-a_{2m}\lambda^{2n-2m}
\end{array}\right),
\label{trunc}\ee
and using it let us define the $n$'th flow of the eigenfunction by
\be
\left( \begin{array}{c}\psi_{1}\\\psi_{2}\end{array}
\right)_{t_{n}}=V^{(n)}(u, \lambda)
\left( \begin{array}{c}\psi_{1}\\\psi_{2}\end{array}
\right).\label{d5}
\ee
Then the compatibility condition of (\ref{a1}) and (\ref{d5}) gives
rise to a zero-curvature representation
\begin{equation}
U_{t_{n}}-V^{(n)}_{x}+[U, V^{(n)}]=0,\quad\quad n=1,2,\cdots.
\end{equation}
Due to the construction of $V^{(n)}$ in (\ref{trunc}) only terms
lowest order in $\lambda$ contribute, yielding the KN hierarchy
\begin{equation}
\left( \begin{array}{c}q\\r\end{array}\right)_{t_{n}}=J
\left (\begin{array}{c}c_{2n-1}\\b_{2n-1}\end{array}\right)
=J\frac {\delta H_{2n-2}}{\delta u}, \quad
J=\left( \begin{array}{cc}0&\partial_{x}\\\partial_{x}&0\end{array}\right),
\label{a2}
\end{equation}
where
\be
H_{2m}=\frac {1}{2m}(4a_{2m+2}-rb_{2m+1}-qc_{2m+1}),\qquad H_{0}=-qr.
\ee

In the above construction all other steps are straightforward, except
the fact that the flow (\ref{a2}) can be written in terms of a
Hamiltonian $H_{2n-2}$, and that the Hamiltonians so obtained are in
involution with respect to the ordinary infinite-dimensional Poisson
bracket \cite{ac85}. [Also in some cases one has to add a lowest order
in $\lambda$ correction term to $V^{(n)}$.] One elegant method to
derive this is by using certain trace identities \cite{gu89}.

\subsection{The constrained flow}
In order to construct a finite dimensional integrable system we will
take $N$ copies of (\ref{a1}) with distinct $\lambda_j$'s
\be
\left( \begin{array}{c}\psi_{1j}\\\psi_{2j}\end{array}\right)_{x}=
U(u, \lambda_{j})
\left( \begin{array}{c}\psi_{1j}\\\psi_{2j}\end{array}\right),
\qquad\quad j=1,...,N,\\
\label{e1}\ee
and these $\psi$'s will be the new dynamical variables (although
sometimes there will be others as well). The additional ingredient we
need is a constraint that relates $u$ to the $\psi$'s. Furthermore
this constraint must be such that it preserves the integrability of
the original system, i.e., it must be invariant under the flows
(\ref{a2}).

One suitable constraint is obtained as follows \cite{z91}.  It is
known \cite{gt93} that for systems (\ref{e1}) with $Tr(U)=0$ we have
(up to a constant factor)
\begin{equation}
\frac {\delta \lambda}{\delta u_{i}}=\frac {1}{2}Tr\left[
\left( \begin{array}{cc} \psi_{1}\psi_{2}& -\psi_{1}^{2}\\
\psi_{2}^{2}&-\psi_{1}\psi_{2}\end{array}\right)\frac {\partial
U(u, \lambda)}{\partial u_{i}}\right],
\end{equation}
which in the present case implies
\begin{equation}
\frac {\delta \lambda}{\delta u}=\frac {1}{2}\left(
\begin{array}{c}\lambda \psi_{2}^{2}\\-\lambda
\psi_{1}^{2}\end{array}\right).
\ee
It is easy to verify that
\be
L\left( \begin{array}{c}\lambda \psi_{2}^{2}\\-\lambda
\psi_{1}^{2}\end{array}\right)=
\la^{2}\left(\begin{array}{c}\lambda \psi_{2}^{2}\\-\lambda
\psi_{1}^{2}\end{array}\right).
\label{a3}
\end{equation}
We then take as our constraint the restriction of the variational
derivatives of conserved quantities $H_{2k_{0}}$ (for any fixed
$k_{0}$) and $\lambda_{j}$ \cite{z94,z91}:
\be
\frac {\delta H_{2k_{0}}}{\delta u}-\beta\sum_{j=1}^{N}
\frac {\delta \lambda_{j}}{\delta u}=0,
\lb{e2}\ee
which in the present case implies
\be
\left( \begin{array}{c}c_{2k_{0}+1}\\b_{2k_{0}+1}\end{array}\right)
-\frac {1}{2}\beta\left( \begin{array}{c}<\Lambda \Psi_{2},
\Psi_{2}>\\-<\Lambda \Psi_{1},\Psi_{1}>\end{array}\right)=0.
\label{a4}\ee
[The constant $\beta$ has been introduced for later convenience.]
Hereafter we denote the inner product in $\bf R^{N}$ by $<.,.>$ and
\begin{equation}
\Psi_1=(\psi_{11},\cdots,\psi_{1N})^{T},\quad\Psi_2=(\psi_{21},
\cdots,\psi_{2N})^{T},\quad \Lambda=diag (\lambda_1,\cdots,
\lambda_N).\end{equation}

It is shown in \cite{z94} that (\ref{e2}) is invariant under all flows
of (\ref{a2}).  The system consisting of (\ref{e1}) and (\ref{e2}) is
called a {\em constrained flow} and can be transformed into a
finite-dimensional integrable Hamiltonian system (FDIHS) by
introducing the so-called Jacobi-Ostrogradsky coordinates.

To deduce the Lax representation for the system (\ref{e1}) and
(\ref{a4}) from the adjoint representation (\ref{d1}), we have to find
the expressions of $a_{m}, b_{m}, c_{m}$ under (\ref{e1}) and
(\ref{a4}). Due to (\ref{d4}), (\ref{a3}) and (\ref{a4}), we may
define the higher order terms \cite{z94} by
\begin{equation}
\left (\begin{array}{c}\tilde c_{2m+1}\\\tilde b_{2m+1}\end{array}\right)
=\frac {1}{2}\beta\left( \begin{array}{c}<\Lambda^{2m-2k_{0}+1}
\Psi_{2},\Psi_{2}>\\-<\Lambda^{2m-2k_{0}+1} \Psi_{1},\Psi_{1}>
\end{array}\right), \qquad m\geq k_{0},
\label{d6}
\end{equation}
and according to (\ref{d3}) and (\ref{e1})
\begin{equation}
\tilde a_{2m}=-\frac {1}{q}(\tilde b_{2m+1}+\frac{1}{2}\tilde b_{2m-1,x})
=\frac {1}{2}\beta<\Lambda^{2m-2k_{0}} \Psi_{1},\Psi_{2}>,
 \qquad m> k_{0}.
\label{d7}
\end{equation}
By using (\ref{d3}), (\ref{d4}), (\ref{e1}) and (\ref{a4}), a direct
calculation gives then expressions for the lower order terms $ a_{2m}$
for $m\leq k_{0}$ and $ b_{2m+1}, c_{2m+1}$ for $m<k_{0}$, which are
denoted also by $\tilde a_{2m}, \tilde b_{2m+1}, \tilde c_{2m+1}$,
respectively.

The construction of $\tilde a_{m}, \tilde b_{m}, \tilde c_{m} $
ensures that under (\ref{e1}) and (\ref{a4})
\begin{equation}
\widetilde V=\sum_{m=0}^{\infty}\left( \begin{array}{cc}
\tilde a_{m}&\tilde b_{m}\\\tilde c_{m}&-\tilde a_{m}
\end{array}\right)\la^{-m}, \label{d8}\end{equation}
satisfies (\ref{d1}) as well. By a direct calculation we find
\begin{equation}\ba{l}
M^{(k_{0})}=\left( \begin{array}{cc}A^{(k_{0})}&B^{(k_{0})}\\
C^{(k_{0})}&-A^{(k_{0})}
\end{array}\right)\equiv\la^{2k_{0}}\widetilde V,\lb{a7}\\
A^{(k_{0})}=\sum_{m=0}^{k_{0}}\tilde a_{2m}\lambda^{2k_{0}-2m}
+\frac{1}{2}\beta\sum_{j=1}^{N}\frac{\la_{j}^{2}\psi_{1j}
\psi_{2j}}{\la^{2}-\la_{j}^{2}},\\
B^{(k_{0})}=\sum_{m=0}^{k_{0}-1}\tilde b_{2m+1}\lambda^{2k_{0}-2m-1}
-\frac{1}{2}\beta\sum_{j=1}^{N}\frac{\la_{j}\la
\psi_{1j}^{2}}{\la^{2}-\la_{j}^{2}},\\
C^{(k_{0})}=\sum_{m=0}^{k_{0}-1}\tilde c_{2m+1}
\lambda^{2k_{0}-2m-1}+\frac{1}{2}\beta\sum_{j=1}^{N}
\frac{\la_{j}\la\psi_{2j}^{2}}{\la^{2}-\la_{j}^{2}}.
\end{array}\ee
Since $\widetilde V$ under (\ref{e1}) and (\ref{a4}) satisfies
(\ref{d1}), the $M^{(k_{0})}$ under (\ref{e1}) and (\ref{a4})
satisfies (\ref{d1}), too, namely
\begin{equation}
M_{x}^{(k_{0})}=[U,M^{(k_{0})}].\label{a6}
\end{equation}
Conversely, the construction of $M^{(k_{0})}$ guarantees that
(\ref{a6}) is just the Lax representation for the system (\ref{e1})
and (\ref{a4}). This can also be verified by a direct calculation.

We present first three systems of (\ref{e1}) and (\ref{a4}) below.

\noindent (a) When $k_{0}=0, \beta=1$, (\ref{a4}) becomes
\be
q=\frac{1}{2}<\La\Psi_{1},\Psi_{1}>, \qquad\quad
r=-\frac{1}{2}<\La\Psi_{2},\Psi_{2}>,
\ee
and then (\ref{e1}) can be written in canonical Hamiltonian form
\be\begin{array}{l}
\Psi_{1x}=-\La^{2}\Psi_{1}+\frac{1}{2}<\La\Psi_{1},\Psi_{1}>
\La\Psi_{2}=\frac{\p \widetilde{H}_{0}}
{\p \Psi_{2}},\\
\Psi_{2x}=-\frac{1}{2}<\La\Psi_{2},\Psi_{2}>\La\Psi_{1}+
\La^{2}\Psi_{2}=-\frac{\p \widetilde{H}_{0}}
{\p \Psi_{1}},\label{a8}\ea\ee
\be
\widetilde{H}_{0}=-<\La^{2}\Psi_{1},\Psi_{2}>+\frac{1}{4}
<\La\Psi_{1},\Psi_{1}><\La\Psi_{2},\Psi_{2}>.
\label{ham0}\ee
The $A^{(0)},B^{(0)},C^{(0)}$ in (\ref{a7}) read
\be \begin{array}{l}
A^{(0)}(\la)=1+\frac {1}{2}\sum_{j=1}^{N}\frac{\la_{j}^{2}
\psi_{1j}\psi_{2j}}{\la^{2}-\la_{j}^{2}},\\
B^{(0)}(\la)=-\frac {1}{2}\la\sum_{j=1}^{N}\frac{\la_{j}
\psi_{1j}^{2}}{\la^{2}-\la_{j}^{2}},\\
C^{(0)}(\la)=\frac {1}{2}\la\sum_{j=1}^{N}\frac{\la_{j}
\psi_{2j}^{2}}{\la^{2}-\la_{j}^{2}}.\label{a9}
\end{array}\ee

\noindent (b) When $k_{0}=1, \beta=\frac{1}{2}$, then (\ref{a4})
gives rise to the constraint
\be q_x=-q^2r-\frac12<\Lambda\Psi_1,\Psi_1>,\quad
r_x=r^2q-\frac12<\Lambda\Psi_2,\Psi_2>,
\label{e4}\ee
and by introducing
\be
q_{1}=q,\qquad p_{1}=r,
\ee
the system (\ref{e1}) and (\ref{e4}) can be written in canonical
Hamiltonian form
\be
\Psi_{1x}=\frac{\p \widetilde{H}_{1}}{\p \Psi_{2}},\qquad
q_{1x}=\frac{\p \widetilde{H}_{1}}{\p p_{1}},\qquad
\Psi_{2x}=-\frac{\p \widetilde{H}_{1}}{\p \Psi_{1}},\qquad
p_{1x}=-\frac{\p \widetilde{H}_{1}}{\p q_{1}},\label{e10} \ee
\be\ba{l}
\widetilde{H}_{1}=-\frac{1}{2}q_{1}^{2}p_{1}^{2}
-<\La^{2}\Psi_{1},\Psi_{2}>
+\frac{1}{2}q_{1}<\La\Psi_{2},\Psi_{2}>-\frac{1}{2}p_{1}
<\La\Psi_{1},\Psi_{1}>.
\end{array}\label{ham1}\ee
The $A^{(1)},B^{(1)},C^{(1)}$ for $M^{(1)}$ are of the form
\be \begin{array}{l}
A^{(1)}(\la)=\la^{2}-\frac{1}{2}q_{1}p_{1}+\frac {1}{4}\sum_{j=1}^{N}
\frac{\la_{j}^{2}\psi_{1j}\psi_{2j}}{\la^{2}-\la_{j}^{2}},\\
B^{(1)}(\la)=-q_{1}\la
-\frac {1}{4}\la\sum_{j=1}^{N}\frac{\la_{j}\psi_{1j}^{2}}{\la^{2}-
\la_{j}^{2}},\\
C^{(1)}(\la)=-p_{1}\la+
\frac {1}{4}\la\sum_{j=1}^{N}\frac{\la_{j}\psi_{2j}^{2}}{\la^{2}-
\la_{j}^{2}}.
\end{array}\label{e11}\ee

\noindent
(c) When $k_{0}=2, \beta=1$, (\ref{a4}) leads to the constraint
\be\ba{rcl}
\frac12 r_{xx}-\frac32 qrr_x+\frac34 q^2r^3&=&-<\Lambda\Psi_2,\Psi_2>,\\
\frac12 q_{xx}+\frac32 qrq_x+\frac34 q^3r^2&=&<\Lambda\Psi_1,\Psi_1>,
\label{e5}\ea\ee
and by introducing the following Jacobi-Ostrogradsky coordinates:
\be
q_{1}=q,\qquad q_{2}=r,\qquad p_{1}=-\frac{3}{16}r^{2}q+
\frac{1}{4}r_{x},\qquad
p_{2}=\frac{3}{16}q^{2}r+\frac{1}{4}q_{x},
\ee
system (\ref{e1}) and (\ref{e5}) can be written in canonical
Hamiltonian form
\be
\Psi_{1x}=\frac{\p \widetilde{H}_{2}}{\p \Psi_{2}},\qquad
q_{ix}=\frac{\p \widetilde{H}_{2}}{\p p_{i}},\qquad
\Psi_{2x}=-\frac{\p \widetilde{H}_{2}}{\p \Psi_{1}},\qquad
p_{ix}=-\frac{\p \widetilde{H}_{2}}{\p q_{i}},\label{a10} \ee
where
\be\ba{rcl}
\widetilde{H}_{2}&=&4p_{1}p_{2}-\frac{3}{4}q_{1}^{2}q_{2}p_{1}
+\frac{3}{4}q_{2}^{2}q_{1}p_{2} -\frac{1}{64}q_{1}^{3}q_{2}^{3}
-<\La^{2}\Psi_{1},\Psi_{2}>\\
&&+\frac{1}{2}q_{1}<\La\Psi_{2},\Psi_{2}>-\frac{1}{2}q_{2}
<\La\Psi_{1},\Psi_{1}>,
\end{array}\label{ham2}\ee
and the $A^{(2)},B^{(2)},C^{(2)}$ for $M^{(2)}$ are of the form
\be \begin{array}{l}
A^{(2)}(\la)=\la^{4}-\frac{1}{2}q_{1}q_{2}\la^{2}+q_{2}p_{2}-
q_{1}p_{1}+\frac {1}{2}\sum_{j=1}^{N}\frac{\la_{j}^{2}
\psi_{1j}\psi_{2j}}{\la^{2}-\la_{j}^{2}},\\
B^{(2)}(\la)=-q_{1}\la^{3}+(\frac{1}{8}q_{2}q_{1}^{2}+2p_{2})
\la-\frac {1}{2}\la\sum_{j=1}^{N}
\frac{\la_{j}\psi_{1j}^{2}}{\la^{2}-\la_{j}^{2}},\\
C^{(2)}(\la)=-q_{2}\la^{3}+(\frac{1}{8}q_{1}q_{2}^{2}-2p_{1})\la+
\frac {1}{2}\la\sum_{j=1}^{N}\frac{\la_{j}\psi_{2j}^{2}}{\la^{2}
-\la_{j}^{2}}.
\end{array}\label{a11}\ee

\section{The main results}
\subsection{The classical Poisson structure}
We now present the classical Poisson structure associated with the Lax
representation for (\ref{a8}), (\ref{e10}) and (\ref{a10}).  With
respect to the standard Poisson bracket, it is found by a direct
calculation that both $A^{(0)},B^{(0)},C^{(0)}$ and
$A^{(1)},B^{(1)},C^{(1)}$ as well as $A^{(2)},B^{(2)},C^{(2)}$ satisfy
the following relations
\be\begin{array}{l}
\{A(\la),A(\mu)\}=\{B(\la),B(\mu)\}=\{C(\la),C(\mu)\}=0,\\
\{A(\la),B(\mu)\}=\frac{\beta\mu}{\mu^{2}-\la^{2}}
(\mu B(\mu)-\la B(\la)),\\
\{A(\la),C(\mu)\}=\frac{\beta\mu}{\mu^{2}-\la^{2}}
(\la C(\la)-\mu C(\mu)),\\
\{B(\la),C(\mu)\}=\frac{2\beta\la\mu}{\mu^{2}-\la^{2}}
(A(\mu)-A(\la)).
\end{array}\lb{a12}\ee

In \cite{o90} it was pointed out, that the integrability of a system
(along with many other useful properties) can be
shown straightforwardly (see Sec.\ \ref{isec}), if the Poisson
structure can be written in the form (we follow the notation of
\cite{skl})
\be
\{M^{(1)}(\al_1),M^{(2)}(\al_2)\}=[r^{(12)}(\al_1,\al_2),M^{(1)}(\al_1)]
-[r^{(21)}(\al_2,\al_1),M^{(2)}(\al_2)].
\label{rdef}\ee
Here the superscripts refer to the vector space on which the matrices
act non-trivially, and $\alpha_i$, $\alpha_j$ are the corresponding
spectral parameters. The equation itself is defined on $V_1\otimes V_2$,
where $V_i$ are identical 2-dimensional vector spaces, so all matrices
are $2^2\times2^2$-dimensional, for example
$M^{(1)}(\al_1)=M(\al_1)\otimes 1$ and $M^{(2)}(\al_2)=1\otimes
M(\al_2)$. From (\ref{rdef}) we can see that the spectral parameters
$\al_i$ are associated with the vector spaces, so it is not necessary
to write them explicitly. Note that the usual permutation matrix $P$
permutes {\em only} the vector spaces: $r^{(21)}(\al_1,\al_2) =
P^{(12)}r^{(12)}(\al_1,\al_2)P^{(12)}$.

For the Poisson brackets (\ref{a12}) one finds that (\ref{rdef}) holds
with
\be r^{(ij)}= \frac{\beta \al_i\al_j}{\al_j^2-\al_i^2}P^{(ij)}-
\frac{\beta \al_i}{\al_j+\al_i}S^{(ij)}, \quad
S^{(ij)}=\frac12(\sigma_0^{(i)}\otimes\sigma_0^{(j)}+
\sigma_3^{(i)}\otimes\sigma_3^{(j)}).  \label{r1}
\ee
(Here the $\sigma_i$'s are the standard Pauli matrices, and the
permutation matrix $P$ is given by $P^{(ij)}=\frac12
\sum_{n=0}^3\sigma_n^{(i)}\otimes\sigma_n^{(j)}$.)  In fact
(\ref{rdef}) hold for all FDIHS obtained from the constrained flows
(\ref{e1}) and (\ref{a4}). The classical Poisson structure
(\ref{rdef}, \ref{r1}) contains all necessary information of the
present system, and is more rich than the Lax representation
\cite{l86}.

It is well know that any Poisson bracket must satisfy the Jacobi identity
\be
\{M^{(1)},\{M^{(2)},M^{(3)}\}\}+
\{M^{(2)},\{M^{(3)},M^{(1)}\}\}+
\{M^{(3)},\{M^{(1)},M^{(2)}\}\}=0.
\label{jac1}\ee
This equation is defined on $V_1\otimes V_2\otimes V_3$ so, e.g.,
$M^{(2)}=1\otimes M(\al_2)\otimes 1$. If we allow for the possibility
that the $r$'s depend on the dynamical variables, then a direct
application of (\ref{rdef}) to (\ref{jac1}) leads to the requirement
\be
[R^{(123)},M^{(1)}]+[R^{(231)},M^{(2)}]+[R^{(312)},M^{(3)}]=0,
\label{jac2}\ee
where
\begin{eqnarray}
R^{(ijk)}&:=&r^{(ijk)}+\{M^{(j)},r^{(ik)}\}-\{M^{(k)},r^{(ij)}\},
\label{Re}\\
r^{(ijk)}&:=&[r^{(ij)},r^{(ik)}]+[r^{(ij)},r^{(jk)}]+[r^{(kj)},r^{(ik)}]
\label{re}.
\end{eqnarray}
If the $r$'s do not depend on dynamical variables, then the Jacobi
identity (\ref{jac2}) should be satisfied by $r^{(ijk)}=0$. This
equation is {\it almost} the classical Yang-Baxter equation, which
would be obtained if we had $r^{(kj)}=-r^{(jk)}$ (in which case the
last term in (\ref{re}) could be written as $[r^{(ik)},r^{(jk)}]$). It
turns out, however, that most of the examples presented here, e.g,
(\ref{r1}), do not have such a antisymmetry, so the index order in
(\ref{re}) is crucial.

For the dynamical $r$-matrices presented in the next section one finds
that $R^{(ijk)}\neq 0$. In \cite{skl} Sklyanin observed that in such
a case the Jacobi identity (\ref{jac2}) can nevertheless be satisfied,
if
\be
R^{(ijk)}=[X^{(ijk)},M^{(j)}]-[X^{(kij)},M^{(k)}],
\label{rcom}\ee
for some matrix $X$. We call this equation the {\em dynamical, classical
Yang-Baxter equation.} [The special case $X^{(ijk)}=X^{(kij)}$ was used
before in \cite{jc94}.] For the examples presented in Sec.\ \ref{dsec}
the Jacobi identity is indeed satisfied due to (\ref{rcom}), where
$X^{(ijk)}\neq X^{(kij)}$.

\subsection{Integrability}\label{isec}
An immediate consequence of (\ref{rdef}) is that
\be
\{M_{1}^{2}(\la), M_{2}^{2}(\mu)\}=[\overline r_{12}(\la,\mu),
M_{1}(\la)]-[\overline r_{21}(\mu,\la), M_{2}(\mu)],\label{a15}
\ee
where \cite{o90}
\be
\overline r_{ij}(\la,\mu)=\sum_{k=0}^{1}\sum_{l=0}^{1}M_{1}^{1-k}
(\la)M_{2}^{1-l}(\mu)r^{(ij)}(\la,\mu)M_{1}^{k}(\la)M_{2}^{l}(\mu).
\ee
Then it follows from (\ref{a15}) immediately that
\be
4\{TrM^{2}(\la), TrM^{2}(\mu)\}=Tr\{M_{1}^{2}(\la),
M_{2}^{2}(\mu)\}=0,\label{a16}
\ee
which ensures the involution property of the integrals of motion
obtained from expanding $M^2$ in powers of $\lambda$.

For system (\ref{a8})
one obtains
\be
TrM^{2}(\la)=(A^{(0)}(\la))^{2}+B^{(0)}(\la)C^{(0)}(\la)=1
+\sum_{j=1}^{N}\frac{F_0^{(j)}}{\la^{2}-\la_{j}^{2}},\ee
where
\begin{eqnarray}
F_0^{(j)}&=&\la_{j}^{2}\psi_{1j}\psi_{2j}
-\frac{1}{4}<\La \Psi_{1}, \Psi_{1}>\la_{j}\psi_{2j}^{2}\nonumber \\
        & &+\frac {1}{4}\sum_{k\neq j}\frac{\la_{j}\la_{k}}{\la_{k}^{2}
-\la_{j}^{2}}
(\la_{j}\psi_{1j}\psi_{2k}-\la_{k}\psi_{1k}\psi_{2j})^{2},\quad j=1,\cdots,N.
\end{eqnarray}
and we have $\widetilde{H}_{0}=-\sum_{i=1}^N F_0^{(j)}$.

For system (\ref{e10}) we find
\be
TrM^{2}(\la)=(A^{(1)}(\la))^{2}+B^{(1)}(\la)C^{(1)}(\la)=\la^{4}-2
\widetilde{H}_{1}
+\frac{1}{2}\sum_{j=1}^{N}\frac{F_1^{(j)}}{\la^{2}-\la_{j}^{2}},\ee
where
\begin{eqnarray}
F_1^{(j)}&=&\la_{j}^{4}\psi_{1j}\psi_{2j}
-\frac{1}{2}q_{1}\la_{j}^{3}\psi_{2j}^{2}+\frac{1}{2}p_{1}\la_{j}^{3}
\psi_{1j}^{2}
-\frac{1}{8}<\La\Psi_{1},\Psi_{1}>\la_{j}\psi_{2j}^{2} \nonumber \\
& &+\frac {1}{8}\sum_{k\neq j}\frac{\la_{j}\la_{k}}{\la_{k}^{2}-\la_{j}^{2}}
(\la_{j}\psi_{1j}\psi_{2k}-\la_{k}\psi_{1k}\psi_{2j})^{2},\quad j=1,\cdots,N.
\end{eqnarray}

For system (\ref{a10}) one gets
\be
TrM^{2}(\la)=(A^{(2)}(\la))^{2}+B^{(2)}(\la)C^{(2)}(\la)=\la^{8}-
\widetilde{H}_{2}\la^{2}+
F_2^{(0)}+\sum_{j=1}^{N}\frac{F_2^{(j)}}{\la^{2}-\la_{j}^{2}},\lb{a17}\ee
where
\begin{eqnarray}
F_2^{(0)}&=&<\La^{4}\Psi_{1},\Psi_{2}>-\frac{1}{2}q_{1}q_{2}<\La^{2}
\Psi_{1},\Psi_{2}>
-\frac{1}{2}q_{1}<\La^{3}\Psi_{2},\Psi_{2}> \nonumber \\
         &  &+\frac{1}{2}q_{2}<\La^{3}\Psi_{1},\Psi_{1}>
              +(p_{2}q_{2}-q_{1}p_{1})^{2}+(p_{2}+\frac{1}{16}
q_{1}^{2}q_{2})<\La\Psi_{2},\Psi_{2}>\nonumber \\
 & &+(p_{1}-\frac{1}{16}q_{2}^{2}q_{1})<\La\Psi_{1},\Psi_{1}>,\nonumber \\
F_2^{(j)}&=&(\la_{j}^{4}-\frac{1}{2}q_{1}q_{2}\la_{j}^{2}
+p_{2}q_{2}-q_{1}p_{1})\la_{j}^{2}\psi_{1j}\psi_{2j}
+(p_{2}-\frac{1}{2}q_{1}\la_{j}^{2}+\frac{1}{16}q_{1}^{2}q_{2})
\la_{j}^{3}\psi_{2j}^{2}\nonumber \\
          & &+(p_{1}+\frac{1}{2}q_{2}\la_{j}^{2}-\frac{1}{16}
q_{2}^{2}q_{1})\la_{j}^{3}\psi_{1j}^{2}
-\frac{1}{4}<\La\Psi_{1},\Psi_{1}>\la_{j}\psi_{2j}^{2} \nonumber \\
          & &+\frac {1}{4}\sum_{k\neq j}\frac{\la_{j}\la_{k}}{
\la_{k}^{2}-\la_{j}^{2}}
(\la_{j}\psi_{1j}\psi_{2k}-\la_{k}\psi_{1k}\psi_{2j})^{2},\quad
\qquad j=1,\cdots,N.\lb{a18}
\end{eqnarray}
Then equation (\ref{a16}) and, for example, (\ref{a17}) guarantees
that the functionally independent integrals of motion
$\widetilde{H}_{2}$ and $F^{(j)}, j=0,1,\cdots,N,$ are in
involution. This shows the integrability of (\ref{a8}), (\ref{e10})
and (\ref{a10}) in the sense of Liouville \cite{vi78}.

\subsection {Two further examples of classical $r$-matrix}
\subsubsection{The KdV hierarchy}
For the KdV hierarchy \cite{mj81}, the eigenvalue problem is of the
form
\be
\left( \begin{array}{c}\psi_{1}\\\psi_{2}\end{array}\right)_{x}=
U(q, \lambda)
\left( \begin{array}{c}\psi_{1}\\\psi_{2}\end{array}\right), \quad
U(q, \lambda)
=\left( \begin{array}{cc}0&1\\-\lambda-q&0\end{array}\right).\label{a1x}
\ee
the second constrained flow with constraint $q=\frac{1}{8}
<\Psi_{1},\Psi_{1}>$ reads \cite{z91}
\be
\Psi_{1x}=\Psi_{2}=\frac{\p \widetilde{H}}{\p \Psi_{2}},\quad
\Psi_{2x}=-\frac{1}{8}<\Psi_{1},\Psi_{1}>\Psi_{1}-\La\Psi_{1}=
-\frac{\p \widetilde{H}}
{\p \Psi_{1}},\label{b1}\ee
with the Hamiltonian
\be
\widetilde{H}=\frac{1}{2}<\La\Psi_{1},\Psi_{1}>+\frac{1}{2}
<\Psi_{2},\Psi_{2}>+\frac{1}{32}<\Psi_{1},\Psi_{1}>^{2}.
\ee
The Lax representation for (\ref{b1}) is given by (\ref{a6})
\be\ba{l}
M(\la)\equiv \left( \begin{array}{cc}A(\la)&B(\la)\\C(\la)&-A(\la)
\end{array}\right)\\
=\left( \begin{array}{cc}0&1\\-\lambda-\frac{1}{16}<\Psi_{1},\Psi_{1}>&0
\end{array}\right)
+\frac {1}{16}\sum_{j=1}^{N}\frac{1}{\la-\la_{j}}
\left( \begin{array}{cc}\psi_{1j}\psi_{2j}&
-\psi_{1j}^{2}\\\psi_{2j}^{2}&-\psi_{1j}\psi_{2j}
\end{array}\right).\label{b2}\ea\ee
and we have
\be\begin{array}{l}
\{A(\la),A(\mu)\}=\{B(\la),B(\mu)\}=0,\\
\{C(\la),C(\mu)\}=\frac{1}{4}(A(\la)-A(\mu)),\\
\{A(\la),B(\mu)\}=\frac{1}{8(\mu-\la)}(B(\mu)-B(\la)),\\
\{A(\la),C(\mu)\}=\frac{1}{8(\mu-\la)}(C(\la)-C(\mu))-\frac{1}{8}B(\la),\\
\{B(\la),C(\mu)\}=\frac{1}{4(\mu-\la)}(A(\mu)-A(\la)).
\end{array}\label{b3}\ee
Then (\ref{b3}) gives rise to the classical Poisson structure
(\ref{rdef}) for the system (\ref{b1}) (in fact for all constrained
flows of KdV hierarchy) with the $r$-matrix given by
\be
r^{(ij)}(\al_i,\al_j)=\frac{1}{8(\al_j-\al_i)}P^{(ij)}+
\frac{1}{8}S^{(ij)},\quad
S^{(ij)}=\sigma_{-}^{(i)}\otimes\sigma_{-}^{(j)}.
\label{b4}\ee
and this $r$ satisfies the classical Yang-Baxter equations of the form
$r^{(ijk)}=0$ (\ref{re}). [In this case $r^{(ij)}\neq -r^{(ji)}$ and
the index order in (\ref{re}) is important.]

\subsubsection{The AKNS hierarchy}
For the AKNS eigenvalue problem \cite{mj81}
\be
\left( \begin{array}{c}\psi_{1}\\\psi_{2}\end{array}\right)_{x}=
U(u, \lambda)
\left( \begin{array}{c}\psi_{1}\\\psi_{2}\end{array}\right), \quad
U(u, \lambda)
=\left( \begin{array}{cc}-\lambda&q\\r&\lambda\end{array}\right).\label{a1y}
\ee
the first constraint is $r=\frac{1}{2}<\Psi_{2},\Psi_{2}>,
q=-\frac{1}{2}<\Psi_{1},\Psi_{1}>$ and the corresponding flow reads
\be\begin{array}{l}
\Psi_{1x}=-\La\Psi_{1}-\frac{1}{2}<\Psi_{1},\Psi_{1}>\Psi_{2}=
\frac{\p \widetilde{H}}
{\p \Psi_{2}},\\
\Psi_{2x}=\frac{1}{2}<\Psi_{2},\Psi_{2}>\Psi_{1}+\La\Psi_{2}=
-\frac{\p \widetilde{H}}
{\p \Psi_{1}},
\label{b6}\ea\ee
where
\be
\widetilde{H}=-<\La \Psi_{1},\Psi_{2}>-\frac{1}{4}<\Psi_{2},\Psi_{2}>
<\Psi_{1},\Psi_{1}>.
\ee
The Lax representation for (\ref{b6}) is given by (\ref{a6}) with \cite{z93b}
\be
\left( \begin{array}{cc}A(\la)&B(\la)\\C(\la)&-A(\la)
\end{array}\right)
=\left( \begin{array}{cc}-1&0\\0&1\end{array}\right)
+\frac {1}{2}\sum_{j=1}^{N}\frac{1}{\la-\la_{j}}
\left( \begin{array}{cc}\psi_{1j}\psi_{2j}&
-\psi_{1j}^{2}\\\psi_{2j}^{2}&-\psi_{1j}\psi_{2j}
\end{array}\right),
\label{b7}\ee
and one gets
\be\begin{array}{l}
\{A(\la),A(\mu)\}=\{B(\la),B(\mu)\}=
\{C(\la),C(\mu)\}=0,\\
\{A(\la),B(\mu)\}=\frac{1}{\mu-\la}(B(\mu)-B(\la)),\\
\{A(\la),C(\mu)\}=\frac{1}{\mu-\la}(C(\la)-C(\mu)),\\
\{B(\la),C(\mu)\}=\frac{2}{\mu-\la}(A(\mu)-A(\la)).
\end{array}\label{b8}\ee
Then (\ref{b8}) gives rise to the classical Poisson structure
(\ref{rdef}) for the system (\ref{b6}) (in fact for all constrained
flows of AKNS hierarchy) with the $r$-matrix given by
\be
r^{(ij)}(\al_i,\al_j)=\frac{1}{\al_j-\al_i}P^{(ij)}.
\label{b9}\ee

\section {Two examples of dynamical $r$-matrix}\label{dsec}
The above examples had an $r$-matrix that depended only on the
spectral parameters. We will now present two restricted flows that
lead to a dynamical $r$-matrix.

\subsection{The TG hierarchy}
Let us first consider the TG hierarchy associated with the following
eigenvalue problem \cite{gt89}
\begin{equation}
\left( \begin{array}{c}\psi_{1}\\\psi_{2}\end{array}\right)_{x}=
U(u, \lambda)
\left( \begin{array}{c}\psi_{1}\\\psi_{2}\end{array}\right), \quad
U(u, \lambda)=\left( \begin{array}{cc}-\lambda+\frac{1}{2}q&r\\r&\lambda-
\frac {1}{2}q\end{array}\right).\label{c8}\end{equation}
The first constrained flow, with constraint $q=(<\Psi_{2},\Psi_{2}>-
<\Psi_{1},\Psi_{1}>)/G$, $r=2G$ reads
\be\begin{array}{l}
\Psi_{1x}=-\La\Psi_{1}+\frac{1}{2G}(<\Psi_{2},\Psi_{2}>-<\Psi_{1},
\Psi_{1}>)\Psi_{1}
+2G\Psi_{2}=\frac{\p \widetilde{H}}{\p \Psi_{2}},\\
\Psi_{2x}=2G\Psi_{1}+\La\Psi_{2}-\frac{1}{2G}(<\Psi_{2},\Psi_{2}>-
<\Psi_{1},\Psi_{1}>)\Psi_{2}
=-\frac{\p \widetilde{H}}{\p \Psi_{1}},\ea
\label{c9}\ee
where
\be
\widetilde{H}=-<\La \Psi_{1},\Psi_{2}>+G(<\Psi_{2},\Psi_{2}>-
<\Psi_{1},\Psi_{1}>), \quad G=\sqrt{<\Psi_{1},\Psi_{2}>}.
\ee
Using the method in \cite{z93a}, we obtain the Lax representation for
(\ref{c9}) given by (\ref{a6}) with
\be
\left( \begin{array}{cc}A(\la)&B(\la)\\C(\la)&-A(\la)
\end{array}\right)
=\left( \begin{array}{cc}-\frac{1}{2}\la&G\\G&\frac{1}{2}\la
\end{array}\right) +\sum_{j=1}^{N}\frac{1}{\la-\la_{j}}
\left( \begin{array}{cc}\psi_{1j}\psi_{2j}&
-\psi_{1j}^{2}\\\psi_{2j}^{2}&-\psi_{1j}\psi_{2j}
\end{array}\right)
.\label{c10}\ee
One gets
\be\begin{array}{l}
\{A(\la),A(\mu)\}=0,\quad \{B(\la),B(\mu)\}=\frac {1}{G}(B(\la)-B(\mu)),\\
\{C(\la),C(\mu)\}=-\frac {1}{G}[C(\la)-C(\mu)],\quad
\{A(\la),B(\mu)\}=\frac{2}{\mu-\la}(B(\mu)-B(\la)),\\
\{A(\la),C(\mu)\}=\frac{2}{\mu-\la}(C(\la)-C(\mu)),\\
\{B(\la),C(\mu)\}=\frac{4}{\mu-\la}(A(\mu)-A(\la))+\frac {1}{G}(B(\la)+C(\mu)),
\end{array}\label{c11}\ee
which gives rise to the classical Poisson structure (\ref{rdef}) for
the system (\ref{c9}) with the dynamical $r$-matrix given by
\be
r^{(ij)}(\al_i,\al_j)=\frac{2}{\al_j-\al_i}P^{(ij)}+\frac {1}{2G}S^{(ij)},\quad
S^{(ij)}=\sigma_{3}^{(i)}\otimes\sigma_{1}^{(j)}.
\ee
This satisfies the classical, dynamical Yang-Baxter equations
(\ref{Re},\ref{rcom}) with
\be
X^{(ijk)}=-\frac{1}{2G^3}\sigma_{3}^{(i)}\otimes\sigma_{3}^{(j)}
\otimes\sigma_{1}^{(k)}.
\ee

\subsection{The Wadati-Konno-Ichikawa hierarchy}
Finally consider the Wadati-Konno-Ichikawa (WKI) hierarchy associated
with the following eigenvalue problem \cite{mw79}
\begin{equation}
\left( \begin{array}{c}\psi_{1}\\\psi_{2}\end{array}\right)_{x}=U(u,
\lambda)
\left( \begin{array}{c}\psi_{1}\\\psi_{2}\end{array}\right), \qquad U(u,
\lambda)
=\left( \begin{array}{cc}\lambda&\lambda q\\\lambda r&-\lambda
\end{array}\right).
\label{c1}\end{equation}
By using the method in \cite{z93a,z94,z91}, we obtain the first
constrained flow, with the constraint $q=-<\La\Psi_{1},\Psi_{1}>/G$,
$r=<\La\Psi_{2},\Psi_{2}>/G$ as follows
\be\begin{array}{l}
\Psi_{1x}=\La\Psi_{1}-\frac{1}{G}<\La\Psi_{1},\Psi_{1}>\La\Psi_{2}=
\frac{\p \widetilde{H}}
{\p \Psi_{2}},\\
\Psi_{2x}=\frac{1}{G}<\La\Psi_{2},\Psi_{2}>\La\Psi_{1}-\La\Psi_{2}=
-\frac{\p \widetilde{H}}{\p \Psi_{1}},\label{c2}
\ea\ee
where
\be
\widetilde{H}=<\La \Psi_{1},\Psi_{2}>-G,\qquad G=\sqrt{1+<\La
\Psi_{1},\Psi_{1}><\La\Psi_{2},\Psi_{2}>}.
\ee
The Lax representation for (\ref{c2}) is given by (\ref{a6}) with
\be
\left( \begin{array}{cc}A(\la)&B(\la)\\C(\la)&-A(\la)
\end{array}\right)
=\frac {1}{2}\left( \begin{array}{cc}G&0\\0&-G\end{array}\right)
+\frac {1}{2}\sum_{j=1}^{N}\frac{\la_{j}}{\la-\la_{j}}
\left( \begin{array}{cc}\la_{j}\psi_{1j}\psi_{2j}&
-\la\psi_{1j}^{2}\\\la\psi_{2j}^{2}&-\la_{j}\psi_{1j}\psi_{2j}
\end{array}\right).\label{c3}\ee
and we have
\be\begin{array}{rcl}
\{B(\la),B(\mu)\}&=&\{C(\la),C(\mu)\}=0,\\
\{A(\la),A(\mu)\}&=&\frac{1}{2G}<\La\Psi_{2},\Psi_{2}>(\la B(\la)-
\mu B(\mu))\\
                  &&+\frac{1}{2G}<\La\Psi_{1},\Psi_{1}>(\la C(\la)-
\mu C(\mu)),\\
\{A(\la),B(\mu)\}&=&\frac{\la\mu}{\la-\mu}B(\la)-\frac{\mu^{2}}{\la-
\mu}B(\mu)+\frac{1}{G}<\La\Psi_{1},\Psi_{1}>\mu A(\mu),\\
\{A(\la),C(\mu)\}&=&\frac{\la\mu}{\mu-\la}C(\la)-\frac{\mu^{2}}{\mu-
\la}C(\mu)+\frac{1}{G}<\La\Psi_{2},\Psi_{2}>\mu A(\mu),\\
\{B(\la),C(\mu)\}&=&\frac{2\la\mu}{\la-\mu}(A(\la)-A(\mu)),
\end{array}\label{c4}\ee
This leads to the classical Poisson structure (\ref{rdef}) for
(\ref{c2}) with the $r$-matrix given by
\be\ba{l}
r^{(ij)}(\al_i,\al_j)=\frac{\al_i\al_j}{\al_j-\al_i}P^{(ij)}-\al_i S^{(ij)}
+\frac{\al_i}{2G}E^{(ij)},\\
S^{(ij)}=\frac{1}{2}(\sigma_{0}^{(i)}\otimes\sigma_{0}^{(j)}
+\sigma_{3}^{(i)}\otimes\sigma_{3}^{(j)}),
\qquad E^{(ij)}=F^{(i)}\otimes\sigma_{3}^{(j)},
\\ F^{(i)}=<\La\Psi_{1},\Psi_{1}>\sigma_{+}^{(i)}
 -<\La\Psi_{2},\Psi_{2}>\sigma_{-}^{(i)}.\ea
\lb{c5}\ee
This $r$-matrix satisfies the dynamical classical Yang-Baxter
equation (\ref{Re},\ref{rcom}), with
\be
X^{(ijk)}=-\frac{\al_i\al_j}{2G^{3}}[F^{(i)}\otimes F^{(j)}\otimes
\sigma_{3}^{(k)}
+2\sigma_{+}^{(i)}\otimes\sigma_{-}^{(j)}\otimes\sigma_{3}^{(k)}
+2\sigma_{-}^{(i)}\otimes\sigma_{+}^{(j)}\otimes\sigma_{3}^{(k)}].
\ee

\section{Conclusions}
In this paper we have discussed the classical Poisson structure and
the related (dynamical) $r$-matrix for some finite dimensional
integrable Hamiltonian systems. These integrable systems were derived
by constraining the integrable flow of an evolution equation in a
particular way \cite{z93a,z94,z91,z93b}.

The possibility of a {\em dynamical} $r$-matrices has been known for
some time now, but no general theory for such systems has been
developed so far.  It is therefore important to derive examples using
different methods, in order to find what the essential features are.
For example, one may ask in what way the Jacobi identities are
satisfied.  The examples presented in Sec.\ \ref{dsec} belong to the
class that satisfy them only through the most general form proposed so
far, Eq. (\ref{rcom}). The method presented in this paper can probably
be used to generate still other types of interesting examples.

\section*{Acknowledgments}
One of us (YBZ) would like to express his gratitude to the Academy of
Finland for financial support. This work was also partially supported
by the Chinese National Basic Research Project 'Nonlinear Science'.

\end{document}